# Staged Laser Wakefield Acceleration for Saturated Lasing of Bandwidth-Tunable Free-Electron Lasers from EUV to X-ray

Hengyuan Xiao[1], Fei Li[2*], Shuang Liu[2], Yuchen Jiang[2,4], Siqin Ding[1], Zhi Song[2], Jianfei Hua[1], Wei Lu[2,3*]

1. Department of Engineering Physics, Tsinghua University, Beijing 100084, China.

2. Institute of High Energy Physics, Chinese Academy of Sciences, Beijing 100049, China.

3. Beijing Academy of Quantum Information Science, Beijing 100193, China.

4. University of Chinese Academy of Sciences, Beijing 100049, China.

\* Corresponding authors: Fei Li (lifei25@ihep.ac.cn); Wei Lu (weilu@ihep.ac.cn)

# Abstract

Free-electron lasers (FELs) provide a revolutionary tool for capturing the structure and dynamics of matter in real time at the atomic scale. The size and cost of FELs can be substantially reduced by using laser wakefield acceleration (LWFA), which offers acceleration gradients orders of magnitude beyond radiofrequency technology, producing multi-GeV electron beams within tens of centimeters. This compactness opens the possibility of integrating multiple operating modes - from the EUV to X-rays including broadband operation – into one facility. Realizing this vision, however, faces key challenges: current LWFA bunches are too short to sustain sufficient radiation slippage, limiting FEL pulse energy at EUV wavelengths, while the large energy spread and emittance make X-ray lasing even more demanding. Here we present a LWFA-driven FEL scheme that addresses these challenges, enabling multi-mode operation spanning different wavelengths and bandwidths within a single facility. The scheme employs staged acceleration to reach multi-GeV energies while preserving beam quality, combined with a dual-chicane beamline that stretches the bunch to mitigate the radiation slippage for EUV FEL and tailors the energy chirp for diverse FEL bandwidth modes. Simulations demonstrate that the scheme can generate high-quality electron beams with energies up to 7 GeV and tunable energy chirp, enabling both FEL saturation from the EUV to X-ray wavelengths and large bandwidth operation with a bandwidth of up to 11%. This work provides a roadmap for compact, multi-mode FELs based on plasma acceleration, and the high-energy, high-quality beams achieved also point toward compact injectors for next-generation storage-ring light sources.

# Introduction

Laser wakefield acceleration (LWFA) has emerged as a revolutionary accelerator technology, capable of producing acceleration gradients three to four orders of magnitude higher than conventional radio-frequency accelerators. This allows for energy gains over 10 GeV within just a few centimeters, offering the potential to dramatically shrink the size and cost of large accelerator facilities such as free electron lasers (FELs)[1–3] and synchrotron light sources based on storage rings[4–6]. This compactness opens the possibility of integrating multiple operating modes — from the EUV to X-rays including broadband operation — into one facility, which would be extremely challenging with RF accelerator. Substantial progress in LWFA has led to the generation of electron beams exhibiting characteristics that closely approach the demands for light source applications, including percent-level energy spread[7–9], sub-micrometer normalized emittance[10,11], femtosecond-scale beam durations[12–14] and 10 kA-level high peak currents[15,16]. These advances have already enabled milestone demonstrations of FEL gain[17–19], thereby stimulating widespread research interest globally.

However, despite these encouraging demonstrations, the development of a fully operational LWFA-based FEL facility that supports multi-mode operation still faces two primary challenges. First, the realization of saturated lasing remains unattainable at EUV and X-ray wavelengths. At EUV wavelengths, the intrinsically femtosecond-scale electron beams from LWFA suffer severely from radiation slippage effects[20,21], where the emitted light outruns the electron bunch within just a few gain lengths. This phenomenon inhibits the FEL interaction from achieving saturation, and thus limits the output pulse energy to 100 nJ level which is several orders of magnitude lower than the hundred-microjoule level necessary for practical applications. At X-ray wavelengths, the primary obstacle is the generation and beam quality preservation of an electron beam that simultaneously possesses high energy (5-10 GeV), ultra-low slice energy spread (~0.1%) and small normalized emittance (~0.5 mm mrad) required for the saturated lasing of X-ray FEL[22,23].

Beyond the goal of attaining saturated output at short-wavelength FELs, large-bandwidth FELs have emerged as a novel operational mode that has attracted considerable interest in recent years, particularly for applications in molecular structural dynamics and X-ray spectroscopy[24–27]. Consequently, one of the significant challenges of LWFA-based FELs is the realization of a tunable bandwidth operating mode, as it would enhance the operational versatility and extend the potential applications.

To address these two challenges, this work proposes a staged LWFA scheme combined with a dual-chicane phase space manipulation system, enabling saturated lasing from the EUV to X-rays as well as large-bandwidth operation within a single facility. This scheme employs a divide-and-conquer strategy by separating the electron beam injection and acceleration processes into two individual phases. This facilitates the production of high-quality electron beams and the attainment of substantial energy gain. Concurrently, the integrated application of two-stage chicanes allows for flexible manipulation of the beam's phase space to accommodate diverse operating modes. For EUV operation, we utilize a chicane to stretch the electron bunch, mitigating slippage and boosting the output pulse energy to the hundreds of μJ range. For X-ray operation, our scheme can generate a 7 GeV beam with exceptional quality, ultimately leading to saturated lasing at 0.26 nm. In terms of the large-bandwidth FEL operation, a large energy chirp of 9.4% can be achieved by precisely tuning the first chicane and the parameters of the energy booster (2$^{nd}$ LWFA), which finally produces FEL radiation at a wavelength of 29 nm with a bandwidth of 11%. Our integrated scheme establishes a viable pathway toward a multi-mode LWFA-driven FEL facility that achieves both high performance and spectral tunability.

In addition to offering a new pathway for LWFA-based FEL, the proposed scheme can also be utilized to drive an LWFA-based storage ring light source. Storage ring light sources typically require multi-GeV electron beams with a relative energy spread at the 0.1% level and geometric emittance of only a few nanometer-radians (nm rad). The electron beams produced by our scheme meet these requirements and therefore show strong potential as injectors for storage ring light sources.

# Results

## Overview of the scheme

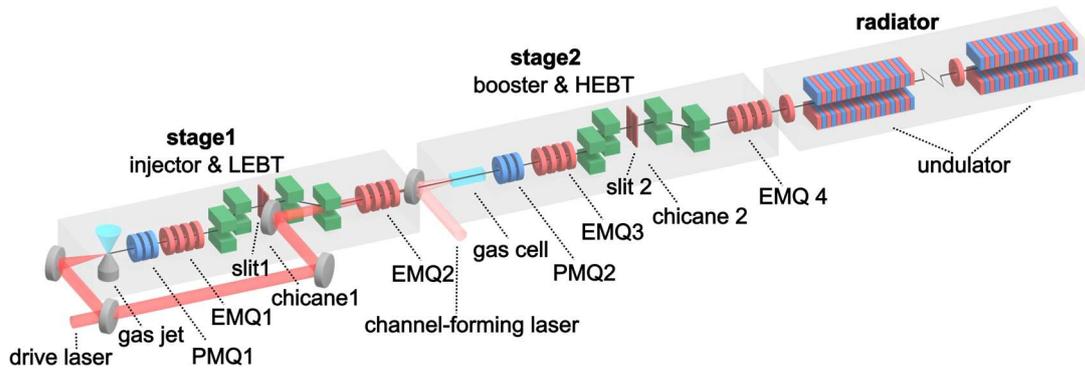

Figure 1. Schematic of the staged LWFA-based FEL scheme combined with a dual-chicane beamline. The undulator beamline is composed of a series of identical sections and only the first and the last sections are shown in the figure.

The layout of the proposed scheme is illustrated in Figure 1 and is organized into two primary stages followed by a radiator. The scheme adopts a divide-and-conquer strategy using an injector-booster configuration to decouple the beam injection from the subsequent acceleration. This design weakens the challenge of simultaneously achieving high beam energy and high beam quality in a single-stage LWFA where the processes of injection and acceleration are usually strongly coupled[7,23,28].

The first stage integrates the injector and the low-energy beam transport (LEBT) line and is dedicated to producing high-quality, moderate-energy electron beams. In the injector section, a joule-class laser interacts with a 6-mm-long supersonic gas jet to generate 500 MeV beams with a high-quality core, which are then captured and manipulated by the LEBT. The LEBT includes a permanent magnet quadrupole assembly (PMQ1) and two electromagnetic quadrupole sets (EMQ1 and EMQ2) to capture the beam and match it to the booster. A chicane (chicane1) equipped with a slit (slit1) within the LEBT selects a portion of the beam with a narrow energy spread (~0.1%) and allows for tuning of the bunch length.

The second stage comprises the booster and the high-energy beam transport (HEBT) line. The booster serves as an energy amplifier to increase the beam energy to the 10 GeV level. A long gas cell (~20 cm) filled with tenuous nitrogen is used as the accelerating medium. To effectively guide the drive laser pulse over such a long distance, a transverse plasma density channel is employed. This channel features a low-density core encircled by a higher-density region and can be formed by the hydrodynamic expansion of the gas[29] induced by the channel-forming laser.

In the energy booster, the desired beam loading effects[30] can be achieved by carefully tuning the laser-plasma parameters, which allows for the precise manipulation of the energy chirp of the electron beams and the ultimate FEL bandwidth. The drive laser pulses for both the booster and the injector are obtained by splitting the same original pulse. This ensures that the accelerated beams and the drive laser exhibit excellent temporal synchronization, rendering their phase jitter negligible. To minimize the emittance growth, tailored plasma density ramps are employed at the entrance and exit of the booster for a matching beam transport at the plasma-vacuum transition.

The layout of the HEBT downstream of the booster resembles that of the LEBT. The beam capturing and matching are conducted by a series of quadrupole assemblies: PMQ2, EMQ3 and EMQ4. Chicane2 with slit2 performs energy selection to reduce the overall energy spread and introduces controlled beam lengthening to mitigate radiation slippage in the EUV FEL.

## Achieving Saturation of EUV FEL by Mitigating Radiation Slippage

A major challenge for LWFA-driven EUV FELs is the radiation slippage effect. The ultrashort, femtosecond-scale electron bunches produced by LWFA typically approach the FEL cooperation length which is defined as the radiation slippage per gain length $L_c = \lambda_r L_g / \lambda_u$. Here, $L_g$ is the gain length, $\lambda_r$ is the radiation wavelength, and $\lambda_u$ is the undulator period. This causes the emitted radiation to outrun the electron bunch and thus prematurely terminate the amplification process long before saturation is reached. This has restricted output energies to the nanojoule (nJ) level[17]. Although schemes employing magnetic chicanes to stretch the bunch have been proposed,[20,21] they were primarily conceived to handle electron beams of relatively poor initial quality. Their effectiveness is often compromised by the beam's large intrinsic energy spread and emittance. Consequently, these approaches have not experimentally demonstrated saturated outputs approaching the hundreds of μJ level.

Our scheme overcomes this limitation by initiating with a high-quality beam and controlling the beam length via chicane2 .

The beam generation process in the injector is depicted in Figure 2a. An ultra-intense 800 nm, 22 fs laser pulse with an energy of 4.2 J and a focal radius $w_0$ of 23 μm is utilized to drive the injector. As the laser pulse enters the underdense plasma with a density of $2.7 \times 10^{18}$ cm$^{-3}$, the ponderomotive force of the laser displaces the ambient electrons, resulting in the formation of a nearly electron-depleted, bubble-like plasma wake (blowout regime[31]) that consists solely of immobile ions, as shown in the left panel of Figure 2a. In order to obtain electron beams with the desired quality, controllable particle injection is achieved by employing the evolution of the plasma wakefield structures induced by the laser self-focusing[32,33]. As the laser pulse is focused to a smaller spot size (see the middle panel of Figure 2a), the wakefield structure expands slowly in the longitudinal direction, and a small portion of the background electrons at the rear of the bubble are captured and accelerated successively. Simulations indicate that this injection process can generate electron beams with the desired excellent quality. Figure 2b shows the longitudinal phase space of the beam core (the blue box in the right panel of Figure 2a). The beam core with a mean current of 25 kA exhibits a slice normalized emittance of 0.2 mm mrad and a slice energy spread of 0.15%.

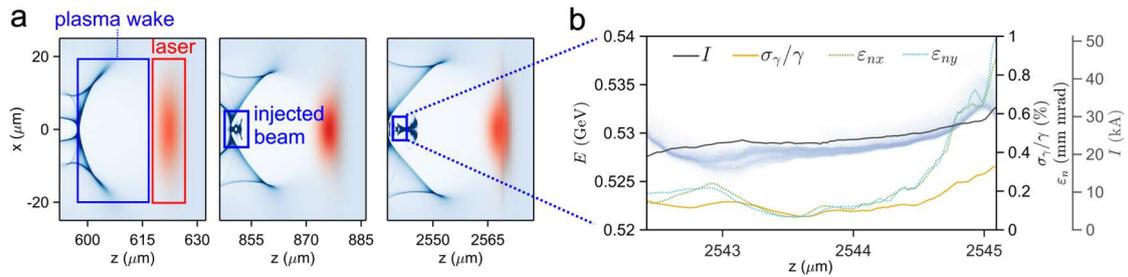

Figure 2. **a**. Simulation results of the electron density (blue) and laser intensity envelope (orange) at three positions in the injector. A nearly electron-depleted, bubble-like plasma wake is generated by the laser plasma interaction, and part of the background electrons are injected and trapped by the accelerating field within the bubble. **b**. The longitudinal phase space and the slice parameters of the core portion of the generated electron beam at the end of the uniform density region (details of the density profile are provided in the Materials and Methods section). This beam portion corresponds to the area enclosed by the blue box in the right panel of **a**.

After the generation of electron beam, the beam core is filtered out by slit1 positioned at the center of chicane1 in the LEBT. The simulation indicates that the LEBT section can select out a 153 pC beam core with a total energy spread of 0.23% from the whole beam. The phase space of the selected beam at the exit of the LEBT is shown in Figure 3a.

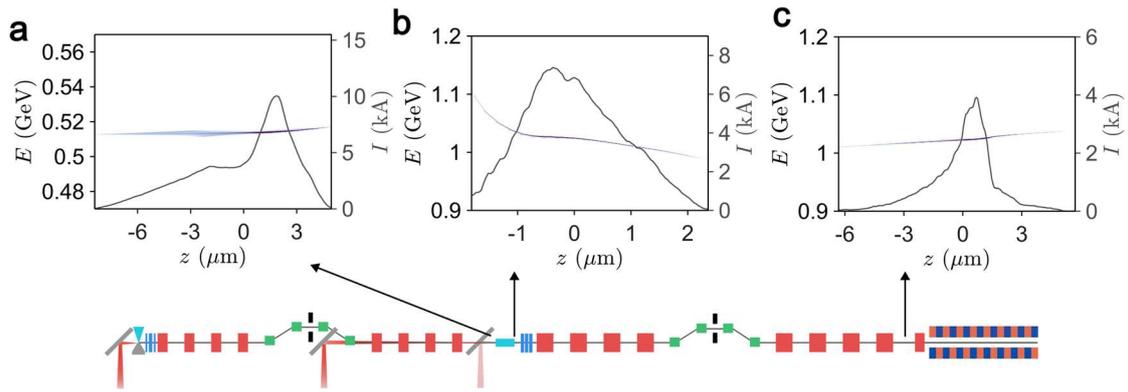

Figure 3. The phase space of the electron beam at the entrance of the booster(**a**), the exit of the booster (**b**) and the entrance of the undulator beamline (**c**) for EUV FEL.

The beam is then injected into the booster stage and accelerated to an energy of 1 GeV, as shown by Figure 3b. Importantly, the beam loading effect is carefully manipulated to yield a near-zero energy chirp at the peak current to generate narrow-bandwidth FEL radiation. After the booster, the beam is stretched by chicane2 in the HEBT (Figure 3 c).

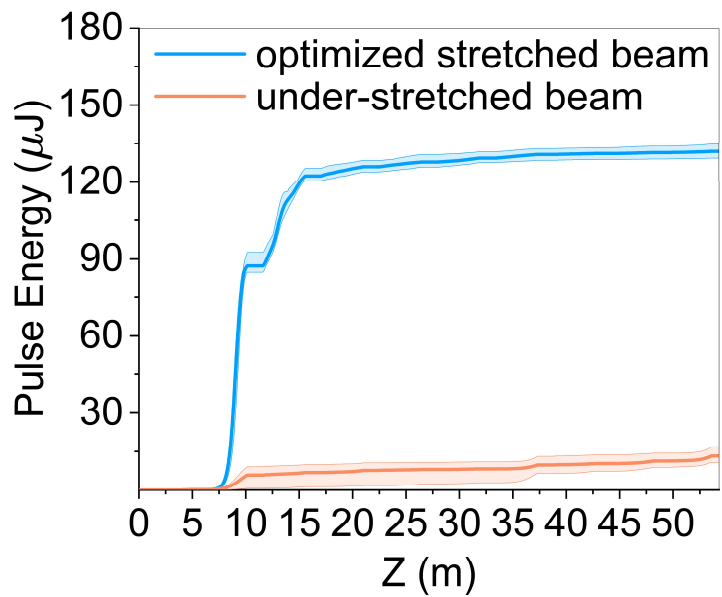

Figure 4. Evolution of the FEL pulse energy along the undulator distance for the optimized stretched and under-stretched beam. The solid line denotes the average over ten simulations with different initial noises, and the shaded region indicates the entire range of these ten results.

To highlight the efficacy of bunch stretching, we compared the FEL performance of the optimized stretched bunch and an under-stretched bunch. Here, the under-stretched case refers to the scenario where the chicane in the HEBT remains active for energy selection but imparts only a small R$_{56}$, leaving the bunch length comparable to the cooperation length. The optimized stretched case corresponds to a chicane strength selected from parameter scans to mitigate radiation slippage while maintaining sufficient peak current for high energy output.

Figure 4 illustrates the evolution of the FEL pulse energy along the undulator beamline for these two cases. The undulator beamline consists of ten undulators with the maximum magnetic field $B_{max} = 1.2$ T and the undulator period $\lambda_u = 3$ cm, which corresponds to the undulator parameter $K \equiv \frac{eB_{max}\lambda_u}{2\pi mc^2} = 2.38$ and radiation wavelength $\lambda_r = 24.8$ nm. Each undulator contains 133 periods.

At the undulator entrance, the under-stretched bunch has an RMS length of 0.38 $\mu$m, approaching the cooperation length of 0.27 $\mu$m at this point. The radiation slippage effect severely hampers the amplification process. The pulse energy fails to grow significantly, remaining trapped at a low level of 10 $\mu$J. When the bunch is optimized stretched to 1.8 $\mu$m, the peak current decreases from 13.7 kA to 4 kA, while the slice energy spread reduces from 0.2% to 0.06%. The final cooperation length remains essentially unchanged at approximately 0.3 $\mu$m, which is only 1/6 of the RMS bunch length at this point. In terms of radiation gain, the optimized stretched beam effectively overcomes the slippage barrier. It exhibits a robust exponential gain starting around Z = 8 m and rapidly reaches saturation. The final stable output pulse energy is approximately 130 $\mu$J. This comparison indicates that the beam stretcher in our beamline design is essential for achieving high-energy saturation in the EUV regime. It should be noted that once the bunch length sufficiently exceeds the cooperation length, the radiation slippage effect is no longer the dominant limiting factor; further increasing the bunch length instead reduces the peak current and thereby weakens the FEL gain. The 1.8 $\mu$m bunch length is therefore selected from our simulation parameter scan as a favorable operating point that balances slippage suppression and peak current preservation.

## Achieving Saturation in X-ray FEL with High-energy High-quality Beams

For lasing at the X-ray wavelengths, the short wavelength renders the effects of slippage negligible. The challenge shifts from beam slippage to the generation of high-energy beams and the preservation of high beam quality.

In an FEL, the radiation wavelength $(\lambda_r)$ varies with the beam energy($\gamma$) according to the resonance condition[22] $\lambda_r = \frac{\lambda_u}{2\gamma^2}(1+\frac{K^2}{2})$. X-ray FEL requires multi-GeV electron beams. After generating a 500 MeV high-quality electron beam using the same method as for the EUV FEL, we inject it into a 20-cm-long gas cell driven by a 600 TW laser with a plasma density of $1.2 \times 10^{17}$ cm$^{-3}$, producing a 7 GeV electron beam. See detailed parameters in Materials and Methods section.

During the acceleration process in the booster, the energy chirp of the electron beam needs to be precisely controlled, and the growth of the emittance must be suppressed as much as possible. Here, we utilized the beam loading effect to achieve the precise energy chirp control for the accelerated beam. The accelerated electron beam with a high current of 8 kA induces significant modifications to the local wakefield structure throughout the acceleration process. This leads to the variations in the slope of the acceleration field, which in turn allows for the control of the energy chirp. By fine tuning the beam current, acceleration phase and laser intensity, we have achieved a region with a constant accelerating field at the beam location (Figure 5a, b) and thus a zero energy chirp at the peak current (Figure 5d) .

When electron beams enter/exit the booster, a dramatic growth of emittance is likely to occur due to the huge disparity between the focusing force in the quadrupole magnet and that in the plasma, unless appropriate beam matching techniques are employed[34,35].

Moreover, if the beam leaves the plasma with a large Twiss parameter $\gamma_T$, the chromatic dispersion in the downstream lattice will also degrade the emittance[36]. Here, we introduce plasma density ramps with a specific gradient to serve as matching sections at both the entrance and exit of the plasma as shown in Figure 5c. At the entrance, the density ramp gradually focuses the rms beam size from 2.5 μm to 0.1 μm, matching the plasma density of the acceleration stage. At the exit, the ramp defocuses the beam and reduces its $\gamma_T$ from 2000 to 34, thereby minimizing the emittance degradation in the HEBT. By optimizing the plasma density ramps, the beam emittance can be well preserved throughout the whole booster stage as shown in Figure 5c.

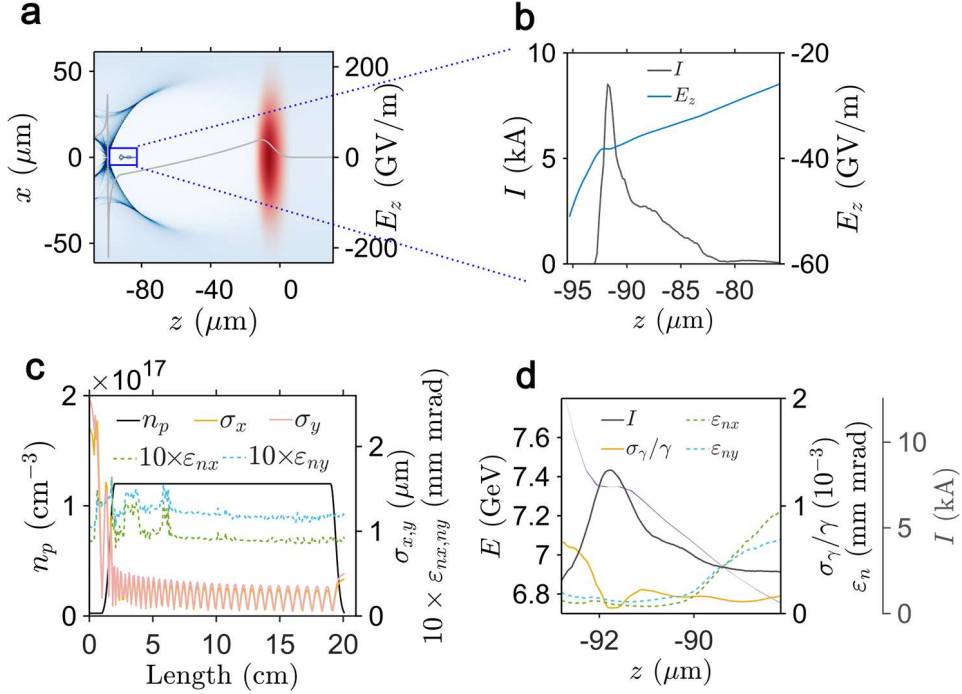

Figure 5. **a**. Simulation results of electron density (blue), laser intensity envelope (orange) and acceleration field (grey line) in the booster for X-ray FEL. **b**. Acceleration field at the location of the electron beam. The slope of the acceleration field is almost zero at the peak current of the electron beam due to the optimized beam loading effect. **c**. Variation of beam sizes, emittances and plasma density in the booster. The beam is focused by the entrance density upramp and defocused by the exit density downramp, while its emittances are well preserved throughout the booster. **d**. The longitudinal phase space and slice parameters of the output electron beam from the booster.

The HEBT uses an approach similar to that of the LEBT to filter the beam energy. The output electron beam of the booster possesses a large energy spread (17%) due to the significant chirp of the beam head, as shown in Figure 5d. The beam head will be filtered out by slit2 in the middle of chicane2. The energy spread of the remaining part is down to 0.12%, meeting the preconditions for narrow-bandwidth FEL lasing.

The undulator beamline configuration remains consistent with that of the EUV FEL, with the exception that $B_{max} = 0.8$ T, corresponding to K = 1.58 and a radiation wavelength of 0.26 nm. The simulations (Figure 6) show that the FEL radiation energy reaches saturation after approximately eight undulators. The saturated output energy of each FEL pulse is 50 µJ, corresponding to $6.5 \times 10^{10}$ photons. The typical spectrum features a central wavelength of 0.256 nm with an rms bandwidth of 0.14%, as shown in Figure 6c.

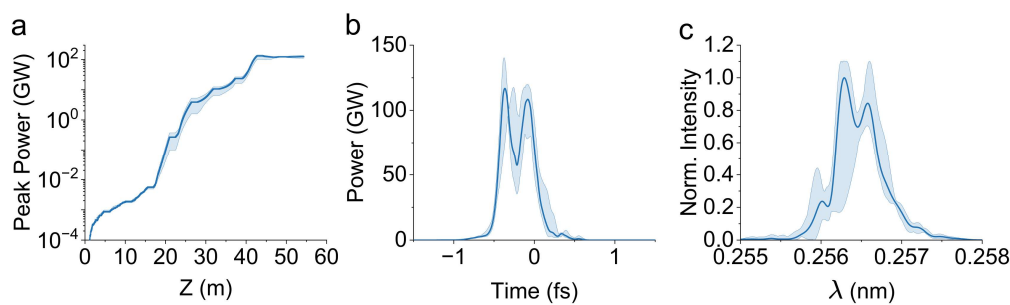

Figure 6. FEL performance of the X-ray FEL. **a**. Gain curve of the peak power along the undulator beamline. **b**. Temporal profile of the FEL radiation power at the exit of undulators **c**. FEL spectrum. The solid line denotes the average over ten simulations with different initial noises, and the shaded region indicates the entire range of these ten results.

# Enabling Large-bandwidth FEL Operation via Chirped Electron Beams

Beyond achieving saturation, another challenge for realizing multi-mode LWFA-driven FEL is to enable large-bandwidth FEL operation which is highly desirable for emerging applications. According to the resonance condition, a large-bandwidth FEL can be generated by an electron beam with a large energy chirp as long as the slice beam emittance and energy spread still reach the criteria for FEL lasing[37]. In this case, different beam slices radiate at different wavelengths in the undulator, thereby broadening the bandwidth of the FEL pulse. In our scheme, this energy chirp is primarily generated by leveraging the beam loading effect. We demonstrate the effectiveness of this mechanism by tracing the phase space evolution of an electron beam that ultimately acquires a large energy chirp, as shown in Figure 7.

By adjusting the chicane1 strength to tune the beam current together with the accelerating phase and the laser intensity in the booster, the beam loading effect can be controlled such that the accelerating field experienced by the beam tail exceeds that at the head (see Figure 7a and b). The slope of the accelerating field results in a significant positive chirp (the energy of the beam tail is higher) of up to 15% at the booster exit (as shown in Figure 7d).

The nonlinear effects in the HEBT can mix the longitudinal phase space. Particles that originate in different slices but exhibit large energy deviations may drift into the same slice at the HEBT exit, thereby degrading the slice quality. Therefore, slit2 in chicane2 is used to remove particles with a large energy deviation and to prevent degradation of the slice quality. The beam entering the undulator (Figure 7e) possesses a charge of 53 pC, an energy chirp of 9.4%, and retains a slice energy spread of 0.1% and a slice emittance of 0.2 mm mrad.

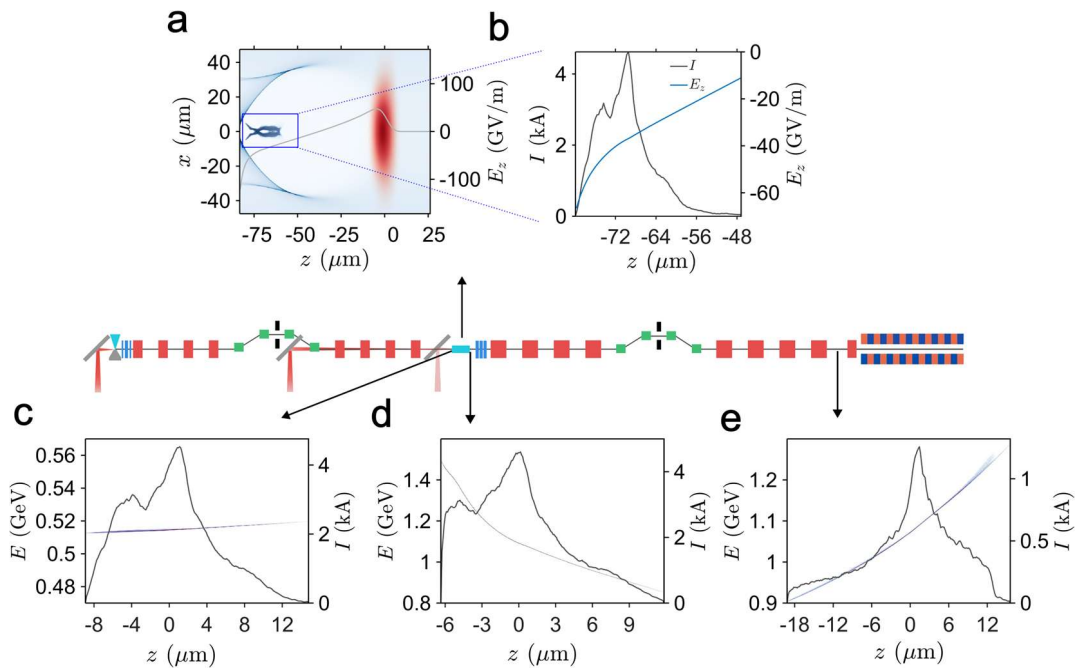

Figure 7. **a**. Simulation results of electron density(blue), laser intensity envelope(orange), and acceleration field (gray line) in the booster for the large-bandwidth FEL. **b**. Acceleration field at the location of the electron beam. **c-e.** The phase space of the electron beam at the entrance of the booster(c), the exit of the booster (d) and the entrance of the undulator beamline (e).

With *K*=2.38, the FEL radiation generated by this electron beam after four undulators is shown in Figure 8. The electron beam at the undulator entrance exhibits significant chirping of 11.8 MeV/μm. Under this large energy chirp, radiation slippage prevents individual FEL slices from remaining in resonance, thereby reducing the output power of the FEL. This can be compensated by introducing a linear taper[21,38], i.e., the undulator parameter *K* varies longitudinally . By matching the local *K* to the local beam energy using a linear taper of $dK/dz = 0.0311 \text{ m}^{-1}$ the FEL resonant condition is satisfied for each beam slice. On average, the FEL pulse reaches saturation after approximately two undulators, delivering 100 μJ of pulse energy (corresponding to $1.3 \times 10^{13}$ photons) with an averaged rms radiation bandwidth of 11%.

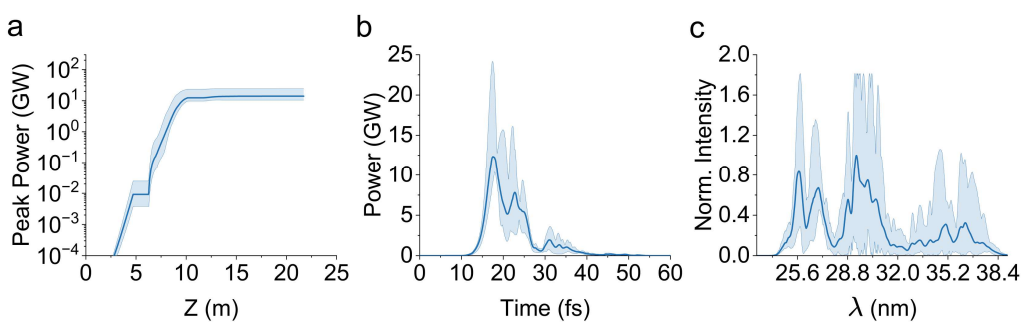

Figure 8. FEL performance of the large-bandwidth FEL. **a**. Evolution of the peak power along the undulator beamline. **b**. Temporal profile of the FEL radiation power at the exit of undulators. **c**. FEL spectrum. The solid line denotes the average over ten simulations with different initial noises, and the shaded region indicates the entire range of these ten results.

# Discussion

Our work proposes a novel staged LWFA scheme based on a dual-chicane structure to enable multi-mode FEL operation within a single facility. Comprehensive start-to-end simulations demonstrate the scheme's capability through three representative operating modes. First, by stretching the bunch, the radiation slippage is mitigated and the EUV FEL pulse energy is boosted by an order of magnitude. Second, the scheme can deliver a high-quality 7 GeV electron beam that drives X-ray FEL radiation to saturation. Third, it can imprint an energy chirp of up to 9.4%, enabling an FEL with an 11% bandwidth to reach saturation.

Although the work in this article only focuses on FEL applications, the proposed scheme can also serve as an injector for storage ring light sources. The advanced storage ring light source based on LWFA is also a current research hotspot internationally. For example, PETRA IV[5], the next-generation synchrotron light source at DESY, requires a beam energy of 6 GeV, an energy spread below 0.3%, and a normalized emittance smaller than 12 mm mrad. Our scheme can generate electron beams with superior beam quality (energy spread 0.12% and emittance 0.2 mm mrad) and comparable beam energy as shown in the section "Achieving Saturation in X-ray FEL with High-energy High-quality Beams". Therefore, our scheme provides a promising solution for the injector of a LWFA-based storage ring light source.

In summary, our study provides a highly feasible, integrated staged LWFA-based scheme for future compact, multi-mode FEL and storage ring light sources. The proposed scheme possesses excellent tunability to accommodate diverse FEL requirements over a broad range of wavelengths and bandwidths, thereby offering a potential pathway toward next-generation light sources.

# Materials and Methods

## LWFA Simulation

The injector is simulated using the fully explicit, relativistic, high-fidelity particle-in-cell (PIC) code OSIRIS[39]. The calculation is conducted in cylindrical geometry with a moving window that propagates at the speed of light. The numerical grid consists of 3600×2500 cells along the radial and longitudinal axes, respectively, with the spatial resolutions of 15.9 nm and 25.5 nm in each direction. The temporal step size is 34 attoseconds. A customized finite difference Maxwell solver[40,41] is used to mitigate the numerical Cherenkov radiation. The plasma density profile consists of a 127 μm density up-ramp, a 2.4 mm uniform density region, and a 3.2 mm density down-ramp.

For the booster stage, the 10 cm level acceleration length makes full PIC simulations computationally expensive. Therefore, we employ QPAD[42], a highly efficient PIC code based on the quasi-static approximation and an azimuthal Fourier-mode decomposition. It retains the critical booster physics while reducing the computational cost by several orders of magnitude. The simulation is carried out in a co-moving window at the speed of light. All parameters used in the simulations presented in the Results section are listed in Table 1.

|  | EUV FEL | X-ray FEL | Large-bandwidth FEL |
|---|---|---|---|
| Plasma parameters | | | |
| Density (cm$^{-3}$) | $1.2\times10^{17}$ | $1.2\times10^{17}$ | $2\times10^{17}$ |
| Total length (mm) | 40 | 201 | 38 |
| Upramp length (mm) | 10 | 10 | 10 |
| Downramp length(mm) | 19 | 11 | 17 |
| Channel depth | 0.38 | 0.38 | 0.37 |
| Laser parameters | | | |
| $a_0$ | 2.8 | 2.8 | 3.4 |
| Pulse width (fs) | 40 | 40 | 31 |
| Size (μm) | 50 | 50 | 39 |
| Power (TW) | 619 | 619 | 576 |
| Distance between beam and laser (μm) | 93 | 91 | 70 |
| Numerical Parameters | | | |
| Grid number | 1024×2048 | 1024×2048 | 512×2048 |
| Grid size (μm) | 0.24×0.14 | 0.24×0.14 | 0.37×0.17 |
| Temporal resolution (fs) | 255.9 | 255.9 | 198.2 |

Table 1. Simulation parameters of the boosters

The density profile of the plasma channel in the booster follows the form $n=n_0(1+\alpha\frac{r^2}{r_0^2})$ where $n_0, r_0$ and $\alpha$ represent the plasma density, laser transverse size and channel depth listed above.

# Beamline Simulation

The LEBT and HEBT sections are simulated using the particle tracking code ELEGANT[43], which accounts for nonlinear beam dynamics and coherent synchrotron radiation (CSR)[44]. In transitioning from the beamline to the LWFA simulations, the full six-dimensional phase space distribution of the electron beam was employed.

The lattice of the LEBT referenced in the Results section is depicted in Figure 9. It consists of a single chicane and eleven quadrupole magnets with a total length of 15.6

m. The first three PMQs are used for the immediate capture of beams at the plasma exit. They are succeeded by eight EMQs. The lengths of the PMQs are 10 cm, 20 cm, and 10 cm, respectively, while both the EMQs and the dipole magnets each have a length of 20 cm. The slit incorporated within the chicane measures 1 cm in width.

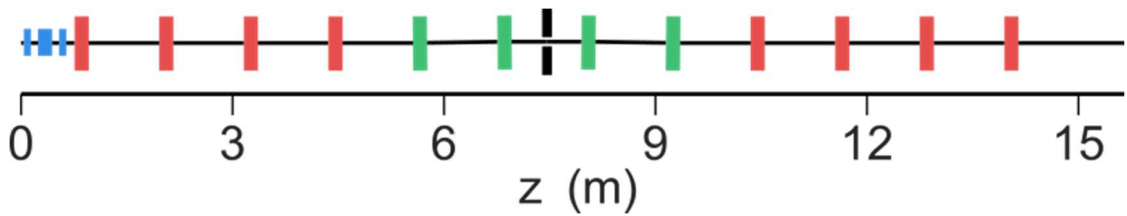

Figure 9. Layout of the LEBT. The blue, red, and green rectangles indicate the PMQs, EMQs and bends, respectively. The black rectangles represent the energy-selection slit.

The HEBT, as described in the Results section, is illustrated in Figure 10 and has a total length of 44.6 m. Its overall configuration is analogous to that of the LEBT; however, the magnets are extended in length to facilitate the transport of electron beams with higher energy. The PMQs are all 20 cm in length, and the EMQs are all 2 m in length. The dipoles are each 1 m long, and the slit is 1 cm in length.

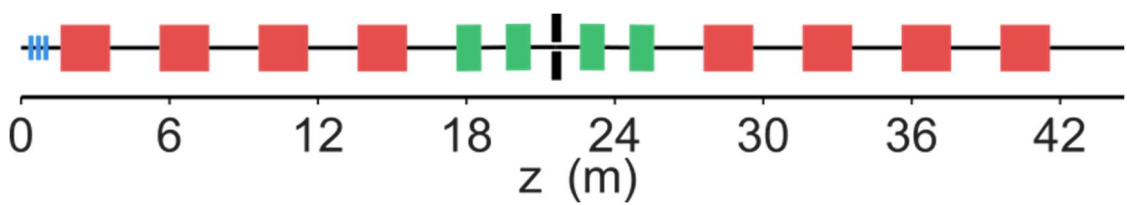

Figure 10. Layout of the HEBT. The blue, red, and green rectangles indicate the PMQs, EMQs and bends, respectively. The black rectangles represent the energy-selection slit.

The strengths of the quadrupoles and dipole magnets and the width of the slit used in the Results section vary from case to case for both the LEBT and the HEBT. Detailed descriptions of the optimization used to determine these parameters are provided in the Optimization of LEBT and the Optimization of HEBT sections.

# FEL Simulation

The dynamics of the free-electron laser are simulated utilizing GENESIS 1.3[45]. The beam transition is modeled using the slice parameters. By employing the slice parameters obtained at the HEBT exit, the simulation reconstructs the beam by distributing macro particles across more refined temporal slices and incorporates the necessary shot noise to initiate the FEL interaction. The undulators used are composed of a series of sections as shown in Figure 11. The quadrupoles each have a length of 0.42 m, and their focusing properties alternate sequentially, i.e., a horizontally focusing (defocusing) quadrupole is followed by a horizontally defocusing (focusing) one. Quadrupoles with the same focusing properties have the same field strength. All drift sections between these elements are 0.51 m in length. Each undulator module contains 133 periods with a total length of 3.99 m. The quadrupole strengths are optimized in conjunction with the HEBT lattice design, and a detailed description is provided in the Optimization of HEBT section.

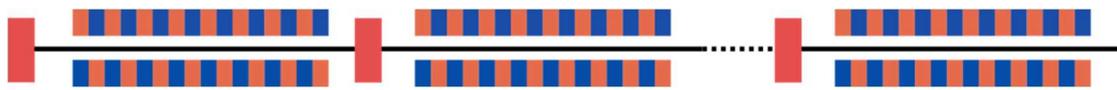

Figure 11. Layout of the undulator beamline. The red rectangles indicate electromagnetic quadrupoles, and the alternating blue and orange blocks correspond to successive undulator poles.

## Optimization of LEBT

The LEBT is divided into two sections and each section is optimized to fulfill its specific function.

The first section, extending from the injector exit to the chicane1 exit, is optimized to ensure that the selected beam exhibits a small energy spread and emittance while preserving sufficient charge. However, nonlinear beam dynamics and collective effects can degrade beam quality and reduce the efficiency of energy selection by disrupting the energy-position correlation at the slit located at the chicane1 midplane. On one hand, the large divergence at the entrance of the LEBT make second-order transport matrix terms non-negligible[36], with $T_{126}$ and $T_{346}$ being the dominant contributors. On the other hand, the combination of a high peak current and ultrashort beam length intensifies the effects of CSR[44,46]. The strengths of the seven quadrupoles in this section are optimized to mitigate the influence of these effects.

The second section extends from the chicane1 exit to the booster entrance. Its optimization focuses on minimizing the transverse beam size at the booster entrance and suppressing the emittance growth arising from nonlinear beam dynamics. The field strengths of the four quadrupole magnets in this section are used as optimization variables.

In both optimizations, the field gradient is limited to 200 T/m for PMQs and 30 T/m for EMQs. The same limits apply to those used in the HEBT and the undulator beamline. Both values are well within the reach of state-of-the-art accelerator technologies.

## Optimization of HEBT

The HEBT is also optimized in two sections. The first section from the booster exit to the chicane2 exit performs the same function as its LEBT counterpart and follows the same optimization strategy as the LEBT.

The second section from the chicane2 exit to the undulator is optimized together with the quadrupoles located between undulators. The optimization purpose is to preserve the slice parameters of the beam and maintain a small beam size inside the undulator to shorten the FEL gain length. The main challenge is to mitigate the growth of the emittances introduced by the second-order transport matrix terms. In addition to the transverse second-order terms mentioned above, longitudinal second-order terms can also degrade the slice quality of the beam by inducing longitudinal phase space mixing.

During optimization, the FEL gain length is evaluated with the Ming Xie formula by inputting the average beam sizes within the undulator and slice parameters at the HEBT exit. The four quadrupoles downstream of the chicane2 exit, together with the strengths of the two groups of quadrupoles between the undulators, are tuned to shorten the gain length.

## Acknowledgment

This work was supported by the Strategic Priority Research Program of the Chinese Academy of Sciences (grant no. XDB0530000), National Natural Science Foundation of China (grant nos.

12375241 and 12305152), and Discipline Construction Foundation of "Double World-class Project". The simulation work is supported by the Center of High Performance Computing, Tsinghua University and Beijing Super Cloud Computing Center. We thank Chao Feng, Zhen Wang and Tao Liu from Shanghai Advanced Research Institute for sharing valuable insights into the current research topic. We thank Bo Peng from Zhengzhou University for useful materials and discussions.

# Author contributions

W.L. conceived and supervised the project. H.X. and W.L. proposed the main idea. H.X., S.L., S.D. and Z.S. carried out the simulations. H.X., F.L., J.H., and W.L. analyzed the results. H.X., Y.J. and F.L. wrote the paper. All the authors contributed extensively to the work presented in this paper.

# Conflict of interest

The authors declare no competing interests.

# Data Availability

The data that support the plots within this paper and other findings of this study are available from the corresponding author upon reasonable request.